# Magnetism and electronic structure of $La_2ZnIrO_6$ and $La_2MgIrO_6$: candidate $J_{eff}$ =1/2 Mott insulators

Guixin Cao,[1,2,6,*] Alaska Subedi,[3,4] S. Calder,[5] J.–Q. Yan,[1,2] Jieyu Yi,[1,6] Zheng Gai,[6] Lekhanath Poudel,[7] David J. Singh,[2] Mark D. Lumsden,[5] A. D. Christianson,[5] Brian C. Sales,[2] David Mandrus [1,2,7,*]

[1]Department of Materials Science and Engineering, University of Tennessee, Knoxville, TN 37996.
[2]Materials Science and Technology Division, Oak Ridge National Laboratory, Oak Ridge, TN 37831.
[3]Max Planck Institute for Solid State Research, Heisenbergstrasse 1, D-70569, Stuttgart, Germany.
[4]Centre de Physique Théorique, École Polytechnique, CNRS, 91128 Palaiseau Cedex, France.
[5]Quantum Condensed Matter Division, Oak Ridge National Laboratory, Oak Ridge, TN 37831.
[6]Center for Nanophase Materials Sciences, Oak Ridge National Laboratory, Oak Ridge, TN 37830.
[7]Department of Physics and Astronomy, University of Tennessee, Knoxville, TN 37996.

**Abstract**

We study experimentally and theoretically the electronic and magnetic properties of two insulating double perovskites that show similar atomic and electronic structure, but different magnetic properties. In magnetization measurements, $La_2ZnIrO_6$ displays weak ferromagnetic behavior below 7.5 K whereas $La_2MgIrO_6$ shows antiferromagnetic behavior (AFM) below $T_N$ = 12 K. Electronic structure calculations find that the weak ferromagnetic behavior observed in $La_2ZnIrO_6$ is in fact due to canted antiferromagnetism. The calculations also predict canted antiferromagnetic behavior in $La_2MgIrO_6$, but intriguingly this was not observed. Neutron diffraction measurements confirm the essentially antiferromagnetic behavior of both systems, but lack the sensitivity to resolve the small (0.22 $\mu_B$/Ir) ferromagnetic component in $La_2ZnIrO_6$. Overall, the results presented here indicate the crucial role of spin-orbit coupling (SOC) and the on-site Coulomb repulsion on the magnetic, transport, and thermodynamic properties of both compounds. The electronic structure calculations show that both compounds, like $Sr_2IrO_4$, are $J_{eff}$ = 1/2 Mott insulators. Our present findings suggest that $La_2ZnIrO_6$ and $La_2MgIrO_6$ provide a new playground to study the interplay between SOC and on-site Coulomb repulsion in a *5d* transition metal oxide.

---
[*] Email: gcao1@utk.edu
[*] Email: dmandrus@utk.edu

**Introduction**

The interplay between spin-orbit coupling (SOC) and on-site Coulomb repulsion in iridates and other *5d* transition metal oxides opens a new field of research in quantum materials [1]. There have been experimental observations of a three-dimensional spin liquid in the hyper-kagome structure of $Na_4Ir_3O_8$ [2], a $J_{eff}$ = 1/2 Mott state in $Sr_2IrO_4$ [1] and $CaIrO_3$ [3], and anomalous "diamagnetism" [4] as well as a dimensionality driven spin-flop transition [5-6] in $Sr_3Ir_2O_7$. Theoretically, many interesting and novel phenomena have been proposed for the iridates, including the Kitaev-Heisenberg model relevant to quantum computing [7], a correlation-enhanced topological insulator in $Na_2IrO_3$ [8], topological Mott insulators in pyrochlore iridates [9], a spin-liquid phase near the Kitaev limit in $A_2IrO_3$ ($A$ = Li, Na) [10], potential high temperature superconductivity in doped $Sr_2IrO_4$ [11], a Weyl semi-metal with Fermi arcs and axion insulators in pyrochlore iridates [12], and the formation of quasi-molecular composite orbitals in $Na_2IrO_3$ [13]. Despite such intriguing studies, the nature of magnetism in the systems with strong SOC remains to be fully elucidated experimentally and theoretically.

In certain iridates, the electronic ground state forms a Mott insulating $J_{eff} = 1/2$ state with a wave function that is a complex linear combination of the $t_{2g}$ orbitals and the spins [14]. These systems may provide another example of a one-band Mott insulating system that bears a familiar resemblance to the high-$T_C$ cuprates [10]. Furthermore, the origin of the intriguing weak ferromagnetic moment in a typical Mott insulating $J_{eff} = 1/2$ system, such as $Sr_2IrO_4$, depends strongly on the Ir-O-Ir bond geometry [7]. The bond-dependent nature of the magnetic interactions leads to an interesting variety of low-energy Hamiltonians, including the isotropic Heisenberg model and the highly anisotropic quantum compass or Kitaev model [7]. To investigate the role of bond geometry in $J_{eff} = 1/2$ systems, investigations of other crystal structures with different local environments and spin-exchange pathways are required, and the nature of their electronic and magnetic structure needs to be revealed.

While $3d$ systems often show Mott insulating states, these are generally rare in $4d$ and $5d$ systems due to the more extended nature of the orbitals in these systems. B-site ordered double perovskites, however, with general formula $A_2BB'O_6$ offer the possibility of forming Mott insulators with $4d$ and $5d$ cations as the distance between the cations is on the order of 5 to 6 Å. Examples of $4d$ Mott insulators are $Ba_2LaRuO_6$ and $Ca_2LaRuO_6$ [15], and for $5d$ systems there are even spin-1/2 materials such as $Sr_2CaReO_6$ [16] and $Ba_2YMoO_6$ [17].

Both $La_2ZnIrO_6$ and $La_2MgIrO_6$ belong to the family of *B*-site ordered double perovskites $A_2BB'O_6$. This family of Ir(IV) mixed oxides was first explored by Galasso and Darby in 1965, and independently by Blasse in the same year [18-19]. The synthesis of $La_2ZnIrO_6$ and the magnetic properties of both $La_2MgIrO_6$ and $La_2ZnIrO_6$ were first studied by Powell, Gore, and Battle in 1993 [20] and subsequently by Currie *et al.* in 1995 [21]. The problem of rationalizing the disparate magnetism of $La_2ZnIrO_6$ and $La_2MgIrO_6$ was first identified in Ref. [20]. The difficulty is that, despite their structural and chemical similarity, $La_2ZnIrO_6$ is weakly ferromagnetic whereas $La_2MgIrO_6$ is antiferromagnetic in magnetization measurements.

In this paper, we report on the magnetism and electronic structure of $La_2ZnIrO_6$ and $La_2MgIrO_6$ and find that these materials, like $Sr_2IrO_4$ are also $J_{eff}$ = 1/2 Mott insulators. Additionally, as the nearest neighbor Ir 5$d$ overlap is smaller due to a larger Ir-Ir distance and the tetragonal distortion of $IrO_6$ octahedra is minimal, the spin-orbit integrated ground state in these compounds is closer to the atomic $J_{eff}$ = 1/2 limit than in previously explored materials. We address the contrasting magnetic behavior of $La_2ZnIrO_6$ and $La_2MgIrO_6$ by experimental measurements of magnetic susceptibility, heat capacity, resistivity, neutron diffraction, and density functional theory calculations. We find that these compounds become insulating at temperatures far above the magnetic transitions. Our calculations show that the combined effect of SOC and on-site Coulomb repulsion results in a $J_{eff}$ = 1/2 Mott insulating state with a "one band" character similar to cuprate superconductors. Interestingly, although the canted antiferromagnetism is well explained by the density functional theory calculations, the collinear antiferromagnetism displayed by $La_2MgIrO_6$ still cannot be rationalized.

**Experimental Details**

Polycrystalline samples of $La_2AIrO_6$ ($A$ = Mg/Zn) were synthesized by solid state reaction of stoichiometric amounts of reactants $La_2O_3$ (99.99%), ZnO/MgO (99.9995%), and $IrO_2$ (99.995%). These mixtures were ground, pelletized, and heated in air at 600 °C overnight, and then heated in air at 1050 °C for one week with intermediate grindings. Similar synthesis procedures were reported in Ref. [22]. Phase purity and crystal structure were determined using both X-ray and neutron diffraction. Additionally, iodometric titration analysis was performed to confirm the Ir oxidation state and oxygen content of the samples following the method reported in Ref. [23]. The $La_2MgIrO_6$ and $La_2ZnIrO_6$ were treated in boiling HBr for 30 minutes, respectively. The elementary bromine was distilled into a KI solution and then the resulting iodine concentration was determined by titration using 0.1 N-$Na_2S_2O_3$ solution. From the results, the oxidation state of Ir was found to be +IV in both samples (+4.00(3) for $La_2MgIrO_6$ and +3.98(1) for $La_2ZnIrO_6$), i.e., the stoichiometries of the samples were $La_2ZnIrO_6$ and $La_2MgIrO_6$, respectively.

Magnetization measurements were performed using a Quantum Design magnetic property measurement system (MPMS) in fields up to 7 Tesla. The specific heat $C(T, H)$ and electrical resistivity $\rho(T, H)$ were performed using a Quantum Design physical property measurement system (PPMS) in applied magnetic fields up to 12 Tesla. Neutron diffraction experiments were performed at the High Flux Isotope Reactor at Oak Ridge National Laboratory, using the HB-1 and HB-3 triple-axis spectrometers, with neutron wavelengths of $\lambda = 2.46$ Å and $\lambda = 2.36$ Å, respectively.

Electronic structure calculations were performed within the local density approximation using the general full-potential linearized augmented plane-wave method as implemented in the ELK software package [24]. We used experimental lattice parameters from [22], but relaxed the internal atomic positions. We also performed some calculations using the lattice parameters and experimental atomic positions that we have experimentally measured, and these calculations give the same physical picture. Muffin-tin radii of 2.2, 2.0, 2.0, 2.0, and 1.6 a.u. for La, Mg, Zn, Ir, and O, respectively, were used. An $8 \times 8 \times 8$ $k$-point grid was used for Brillouin zone integration. The spin-orbit coupling was treated using a second-variational method, and the fully localized limit was used to take into account the double counting in the LDA+$U$ calculations.

**Results and Discussion**

Powder X-ray diffraction (XRD) data of polycrystalline $La_2MgIrO_6$ and $La_2ZnIrO_6$ collected using a PANalytical X'pert PRO MPD at room temperature using Cu $K_{\alpha 1}$ radiation are shown in Fig. 1 (a) and (b). All the lines in the XRD pattern could be indexed to the monoclinic $P2_1/n$ structure. This monoclinic double perovskite structure is derived from the perovskite structure by alternatingly placing Mg/Zn and Ir at the $B$ site such that Mg/Zn and Ir ions each form an fcc lattice, as shown in the right panels of Fig.1. The structural motifs in these compounds are thus

Mg/ZnO$_6$ and IrO$_6$ octahedra arranged in an fcc lattice, with La ions at the *A* site providing charge balance. This is in contrast to previous $J_{eff}$=1/2 materials Sr$_2$IrO$_4$, CaIrO$_3$, and Na$_2$IrO$_3$, where IrO$_6$ octahedra arranged on a two-dimensional plane are the main structural units. Also in contrast to Sr$_2$IrO$_4$, CaIrO$_3$, and Na$_2$IrO$_3$, the IrO$_6$ octahedra in La$_2$MgIrO$_6$ and La$_2$ZnIrO$_6$ have very minimal tetragonal distortions with the Ir-O distances within an octahedron varying by less than 0.5%. The main distortions from the ideal double pervoskite structure in La$_2$MgIrO$_6$ and La$_2$ZnIrO$_6$ are the rotations of the octahedra to reduce the volume by shortening the La-O distances. The octahedra in La$_2$MgIrO$_6$ and La$_2$ZnIrO$_6$ are rotated around the *b* and *c* axes, in contrast to the previously studied $J_{eff}$=1/2 Mott insulators Sr$_2$IrO$_4$ and CaIrO$_3$ that are only rotated around one axis. The Rietveld refinement (using GSAS with $R_p \sim$ 6.8, $R_{wp} \sim$ 10.1 and $\chi^2 \sim$ 1.1 for La$_2$MgIrO$_6$ and with $R_p \sim$7.1, $R_{wp} \sim$ 9.8 and $\chi^2 \sim$ 1.3 for La$_2$ZnIrO$_6$) of the XRD pattern with Cu K$\alpha_1$ radiation (1.54059 Å) yielded the crystallographic data as listed in Table I.

The dc magnetic susceptibility measured in an applied magnetic field of $\mu_0 H$ = 1 Tesla shows a weak ferromagnetic transition (actually, this is a canted antiferromagnetic (CAF) transition as determined from neutron diffraction) near 7.5 K for La$_2$ZnIrO$_6$ and an antiferromagnetic transition near 12 K for La$_2$MgIrO$_6$, as displayed in Fig. 2 (a) and (c). A Curie-Weiss fit of the high-temperature data of $\chi$ for 50 < T < 350 K yields an effective moment $\mu_{eff}$ of 1.71 $\mu_B$ and a negative Curie-Weiss temperature $\Theta_{CW}$ of -24.0 K for La$_2$MgIrO$_6$ and $\mu_{eff}$ of 1.42 $\mu_B$ and $\Theta_{CW}$ of -3.1 K for La$_2$ZnIrO$_6$ (see Figs. 2(b) and 2(d)). The Curie-Weiss fit for La$_2$ZnIrO$_6$ can hold even to 10 K with similar $\mu_{eff}$ and $\Theta_{CW}$. These effective moments are comparable with previously reported data [21] and other *5d* S=1/2 systems such as Sr$_2$CaReO$_6$ ($\mu_{eff}$ = 1.66 $\mu_B$ and $\Theta_{CW}$ = -443 K) and Ba$_2$YMoO$_6$ ($\mu_{eff}$ = 1.42 $\mu_B$ and $\Theta_{CW}$ = -45 K) [16-17, 25]. The similarity of the magnitude of the fitted effective moments of the two compounds reflects the small influence of the difference between Mg and Zn on the character of the Ir-O bonding.

Fig. 3(a) and 3(b) shows isothermal magnetization *M* versus magnetic field *H*

measured at T = 2 K for $La_2MgIrO_6$ and $La_2ZnIrO_6$, respectively. It can be seen that the *M(H)* curve is proportional to *H* indicating the absence of any ferromagnetic signature in $La_2MgIrO_6$. However, there exists a clear ferromagnetic component for $La_2ZnIrO_6$ as shown in Fig. 3(b). Note that full saturation is not achieved for the ferromagnetic component in $La_2ZnIrO_6$, which is consistent with a CAF ground state.

To further characterize the ferromagnetism in $La_2ZnIrO_6$, we carried out an Arrott analysis to determine the $T_C$ and ordered moment of this compound. The analysis of the spontaneous magnetization (eliminating domain effects) $M_s$ ($M_s \equiv M(H=0)$) and the initial susceptibility $\chi_0$ ($\chi_0 \equiv \partial M/\partial H |_{H=0}$) is performed based on the *M(H)* data measured. The modified Arrott plot [26-27] was used to obtain $T_C$, as shown in Fig. 3 (c), which shows $(M)^{1/\beta}$ vs $(H/M)^{1/\gamma}$ plots. The isothermal curve at 7.5 K is very linear, suggesting the correct $T_C \approx$ 7.5 K. $\beta$ = 0.74 (obey $M_s \sim t^\beta$ for $T < T_C$) and $\gamma$ = 1.10 (obey $\chi_0 \sim t^{-\gamma}$ for $T > T_C$) were also obtained. $M_s$ at 2 K is determined as 0.22 $\mu_B$/Ir from the linear extrapolation of the straight line in the modified Arrott plots with the $M^{1/\beta}$ axis. The 0.22$\mu_B$/Ir moment is smaller than the calculated value of 0.47$\mu_B$/Ir (as discussed below) for the ferromagnetic component of the moment in $La_2ZnIrO_6$, which is probably due to the averaging of the ferromagnetic component of the moment for different directions in polycrystalline samples. [28-30].

Heat capacity $C_p$ measurements for $La_2MgIrO_6$ and $La_2ZnIrO_6$ between 1.9 and 30 K are presented in Fig. 4. It can be seen that there is obvious transition at 11 K for $La_2MgIrO_6$ and at 7 K for $La_2ZnIrO_6$ in the $C_p(T)$ data at $\mu_0H$ =0 T, which is near the AFM transition for $La_2MgIrO_6$ and CAF transition for $La_2ZnIrO_6$, respectively. The 11 K transition in the specific heat for $La_2MgIrO_6$ shows a slight shift to a lower temperature under $\mu_0H$ =12 T, consistent with expectations for an AFM transition. For $La_2ZnIrO_6$, the peak is depressed and moves slightly higher in temperature as expected for a CAF.

Fig.4 (b) and (d) shows difference heat capacity $\Delta C_p = C(T) - C_{lattice}(T)$ and the integrated entropy S(T) = ∫$C_p$/T dT. Reference compounds used to estimate the lattice contribution were $La_2ZnTiO_6$ and $La_2MgTiO_6$. These reference materials were not perfect, as can be seen in Fig. 4 at higher temperatures where the difference in specific heat approaches 1 J/mol-K.

Comparing the $\Delta C_p$ data of both compounds, it can be seen that the transition peak at 7 K in $La_2ZnIrO_6$ is sharper than the peak at 11 K in $La_2MgIrO_6$. This could indicate competing interactions in $La_2MgIrO_6$ that are not fully accounted for in the first principles calculations and may provide a hint as to the cause of the different behavior of the two compounds. The entropy removed by the magnetic transition is S ≈ 3.4 J/mole K in $La_2MgIrO_6$ and S ≈ 3.7 J/mole K in $La_2ZnIrO_6$, which is about 59 % (for $La_2MgIrO_6$) and 64 % (for $La_2ZnIrO_6$) of the value Rln(2) = 5.76 J/mole K expected for ordering of J = 1/2 moments.

Neutron diffraction measurements were performed on $La_2MgIrO_6$ and $La_2ZnIrO_6$ at temperatures from 5-16 K and 2-12 K, respectively. At 4 K both compounds showed additional scattering at Q = 0.79 Å$^{-1}$, indicative of long range antiferromagnetic order. The additional reflection is commensurate with the nuclear structure and is compatible with a k = (0,0,0) propagation vector, which indicates that the antiferromagnetic ordering is describable within a single crystallographic unit cell. The results are shown in Fig. 5, along with magnetic structures predicted from first principles as explained below. Note that $La_2MgIrO_6$ is predicted to be a canted antiferromagnet, like $La_2ZnIrO_6$, whereas magnetization measurements show that $La_2MgIrO_6$ is not canted.

The long-range magnetic ordering temperature observed in neutron scattering is consistent with the onset of magnetic ordering observed in magnetization measurements. As Ir is a strong absorber of neutrons and has a large magnetic intensity reduction with increasing |Q| due to the Ir magnetic form factor, the amount of information that could be obtained from powder diffraction is limited. Although we could confirm the basic antiferromagnetism in both compounds, we could not refine the magnetic structure or compare the details of magnetic order in the two compounds.

It is interesting to compare the magnetism in $La_2MgIrO_6$ and $La_2ZnIrO_6$ with other spin-1/2 5*d* double perovskites with only one magnetic ion at the *B* site. Ferromagnetism has been previously observed in $5d^1$ double perovskite $Ba_2NaOsO_6$,[31] but isostructural and isovalent $Ba_2LiOsO_6$ instead orders antiferromagnetically. [32] Yet another magnetic ground state is found in the S = 1/2 5*d* double perovskite compounds $Sr_2CaReO_6$ and $Sr_2MgReO_6$, which show spin glass behavior in spite of the minimal *B*-site disorder. [33] The wide variety of behavior observed in these nominally similar materials underscores the delicate nature of the interactions in these complex oxides.

The electrical resistivity $\rho$ between *T* = 80 and 350 K for both compounds increases dramatically upon cooling as shown in Fig. 6. As the materials are good insulators well above the magnetic ordering temperatures, this suggests that both are Mott insulators. The $\rho$-*T* curves follow an activation law $\rho(T) \sim exp(\frac{\Delta}{2k_BT})$ (*Δ* is the activation energy and $k_B$ is the Boltzmann constant) with $La_2MgIrO_6$ exhibiting a value of *Δ* ~ 0.16 eV. In $La_2ZnIrO_6$ the $\rho$-*T* curve can be parameterized by two distinct values of *Δ* in regions that correspond to I and II, respectively, as shown in the inset of Fig.6 (b). The *Δ* is inferred to be 0.13 eV in region I and 81.7 meV in region II. Application of a $\mu_0H$ = 12 T magnetic field does not appreciably affect the resisitivity as shown in the Figure.

Our electronic structure calculations for $La_2MgIrO_6$ and $La_2ZnIrO_6$ support the Mott insulating state for these compounds. Table II gives the details of the $\langle \vec{L} \rangle$ and $\langle \vec{S} \rangle$ expectation values computed over Ir muffin-tin spheres and the band gap $E_{gap}$ for some values of *U* and Hund's coupling *J*. The paramagnetic LDA+SOC band structures of $La_2MgIrO_6$ and $La_2ZnIrO_6$ are shown in Fig 7 (a1) and (b1), respectively. In both compounds, the states near the Fermi level are dominated by the Ir $t_{2g}$ bands. These six spin degenerate $t_{2g}$ bands arising from two Ir atoms lie between -1.0 and 0.3 eV relative to the Fermi level. The Ir $t_{2g}$ bands have a narrow band width of ~1.3 eV due to the large Ir-Ir separation in the double perovskite structure, which results in smaller inter-site hopping. As one expects for crystal field-split electronic states, the Ir $t_{2g}$ bands also show O $p$ character, and the O $p$ bands that lie between -7.5 and -1.7 eV (not shown) also correspondingly show Ir $d$ character. As in the case of $Sr_2IrO_4$ [14, 34-36] and $CaIrO_3$ [37], SOC splits the Ir $t_{2g}$ bands of $La_2MgIrO_6$ and $La_2ZnIrO_6$ into a lower-lying group of four and a higher-lying group of two spin-degenerate bands, which would correspond to effective total angular momenta $J_{eff}$ = 3/2 and 1/2, respectively, in the atomic and strong SOC limit. The $J_{eff}$ = 3/2 bands are completely filled while the $J_{eff}$ = 1/2 bands are half-filled, consistent with the nominal ionic state of $Ir^{4+}$ ($5d^5$). Since the $J_{eff}$ = 1/2 bands are half-filled, $La_2MgIrO_6$ and $La_2ZnIrO_6$ are metallic within the LDA, which is contrary to the Mott insulating behavior that we have observed experimentally. This suggests that the on-site Coulomb repulsion plays a significant role in the electronic structure of these compounds. Indeed, the $J_{eff}$ = 1/2 bands have a narrow width of ~0.5 eV, and we find that a value for on-site Coulomb repulsion $U$ = 1.0 eV yields insulating states for $La_2MgIrO_6$ and $La_2ZnIrO_6$ by splitting the $J_{eff}$ = 1/2 bands into fully occupied lower Hubbard bands and unoccupied upper Hubbard bands (see Figs. 7(a2) and (b2)). The combination of spin orbit and Coulomb correlation to obtain an insulating state in $5d^1$ double perovskite was also discussed by Lee and Pickett in the context of $Ba_2NaOsO_6$. [38] This is a Mott insulating state in the sense that a single-particle theory such as standard approximate density functional theories that are implemented using Kohn-Sham formalism (e.g. the local density approximation) cannot explain the insulating state, and an explicit treatment of on-site Coulomb repulsion is needed.

The LDA+SO+$U$ calculations for the ground states of $La_2MgIrO_6$ and $La_2ZnIrO_6$ give canted antiferromagnetic orderings for the unit cells used in our calculations. (Note that our unit cell has two Ir ions, one at (0, 0.5, 0) and another at (0.5, 0, 0.5). They are not within the same $a$-$b$ plane.) For $U$ = 1.0 eV, the calculations for $La_2MgIrO_6$ give a total moment of 0.52 $\mu_B$/Ir, with an orbital moment of 0.32 $\mu_B$/Ir and spin moment of 0.20 $\mu_B$/Ir and for $La_2ZnIrO_6$ a total moment of 0.55 $\mu_B$/Ir with an orbital moment of 0.33 $\mu_B$/Ir and spin moment of 0.23 $\mu_B$/Ir. In both compounds, the orbital and spin moments are nearly parallel to each other, as expected for $J_{eff}$ = 1/2 moments. (See Table II for the $\langle \vec{L} \rangle$ and $\langle \vec{S} \rangle$ expectation values over Ir muffin-tin spheres.) The total moments are canted along the $b$ axis such that there is a net ferromagnetic moment of 0.33 $\mu_B$/Ir in La$_2$MgIrO$_6$ and 0.47 $\mu_B$/Ir in La$_2$ZnIrO$_6$. This is consistent with the rotation of the two IrO$_6$ octahedra in our unit cell, which are canted towards each other in the $ab$ plane. On the other hand, the octahedra are tilted in the same direction along $c$ axis. As expected, we do not find any net ferromagnetic moment along the $c$ direction.

**Conclusions**

Our calculations give orbital and spin moments that are different from what is expected for the ideal $J_{eff}$ = 1/2 case. (In the ideal case, the orbital moment is 0.67 $\mu_B$ and spin moment is 0.33 $\mu_B$.) A major factor for this deviation should be the covalency between Ir $5d$ and O $2p$ states, although, this should proportionally reduce both the orbital and spin moments. The paramagnetic LDA DOS (Fig. 8 (a1) and (b1)) shows that the projections onto $d_{x^2-y^2}$, $d_{yz}$ and $d_{zx}$ orbitals that form the $t_{2g}$ states in these compounds are almost degenerate, which indicates that the deviation from the ideal $J_{eff}$ = 1/2 case is not due to the lifting of the degeneracy of the $t_{2g}$ states that might arise from the very small distortion of the IrO$_6$ octahedra. Furthermore, we find some contribution of $d_{xy}$ and $d_{z^2}$ (which form the $e_g$ states) orbitals in the $t_{2g}$ manifold of both compounds, and this mixing of the $e_g$ states that have effective orbital moments of zero should reduce the orbital moment, but might enhance the spin contribution. This $t_{2g} - e_g$ mixing is allowed due to the octahedral tilts and therefore is a reflection of a band formation that is sensitive to these tilts. In addition, the paramagnetic LDA+SO DOS (Fig. 8 (a2) and (b2) ) shows that there is some mixing between $J_{eff}$ = 1/2 and 3/2 states due to inter-site hopping in both compounds, and this will further move these systems away from the ideal $J_{eff}$ = 1/2 case.

Although the canted antiferromagnetism found in the calculation could only be detected in $La_2ZnIrO_6$, the peculiar electronic and magnetic properties of both $La_2MgIrO_6$ and $La_2ZnIrO_6$ can be understood as characteristics of $J_{eff}=1/2$ Mott insulators that are closer to the atomic limit than $Sr_2IrO_4$ and $CaIrO_3$. The residual ferromagnetism is manifest in the experiments only for $La_2ZnIrO_6$ presumably due to its larger canting angle. Local moments are larger in these compounds than in $Sr_2IrO_4$, but still away from the ideal $J_{eff}=1/2$ case due to the Ir $5d$ – O $2p$ covalency, presence of some $e_g$ states in the $t_{2g}$ manifold due to the octahedral rotation, and mixing of $J_{eff} = 1/2$ and $3/2$ states due to inter-site hopping. In particular, these two compounds show that the moments are away from the ideal case of the strong SOC limit even when the tetragonal distortion of the $IrO_6$ is minimal and the inter-site hopping between $Ir^{4+}$ ions is reduced. Previously, it was found that the Mott insulating state in $CaIrO_3$ is also further from the ideal $J_{eff} = 1/2$ state [37,39], but the presence of tetragonal distortion of the $IrO_6$ in $CaIrO_3$ complicates the analysis of the role of inter-site hopping in modifying the ideal $J_{eff} = 1/2$ state.

In conclusion, the weakly ferromagnetic behavior found in magnetization with anomalous 0.22 μ$_B$/Ir below $T_c$ < 7.5 K in $La_2ZnIrO_6$ originates from canted antiferromagnetism. Our neutron scattering results reveal long range antiferromagnetic ordering and agree well with the DFT calculations. From the DFT calculations, we find that the Ir $t_{2g}$ bands in both compounds are split into fully filled $J_{eff} = 3/2$ and half-filled $J_{eff} = 1/2$ bands. The Ir $5d$ bands are narrower than in $Sr_2IrO_4$ and $CaIrO_3$, and in this sense $La_2MgIrO_6$ and $La_2ZnIrO_6$ are closer to the atomic limit than $Sr_2IrO_4$ and $CaIrO_3$. The inclusion of a modest on-site Coulomb repulsion further splits the half-filled $J_{eff} = 1/2$ and opens a narrow gap, leading to a Mott insulating state. Our present findings suggest that $La_2ZnIrO_6$ and $La_2MgIrO_6$ are spin-orbit integrated Mott insulators that can provide a new playground for the study of novel behavior in the vicinity of the $J_{eff} = 1/2$ state.

**Acknowledgments**

We acknowledge useful discussion with Satoshi Okamoto and George Jackeli.

**References**

[1] B. J. Kim, H. Ohsumi, T. Komesu, S. Sakai, T. Morita, H. Takagi, and T. Arima, Science **323**, 1329 (2009).

[2] Y. Okamoto, M. Nohara, H. Aruga-Katori, and H. Takagi, Phys. Rev. Lett. **99**, 137207 (2007).

[3] K. Ohgushi, J.-I. Yamaura, H. Ohsumi, K. Sugimoto, S. Takeshita, A. Tokuda, H. Takagi, and T.-H. Arima, arXiv:1108.4523.

[4] G. Cao, Y. Xin, C. S. Alexander, J. E. Crow, P. Schlottmann, M. K. Crawford, R. L. Harlow, and W. Marshall, Phys. Rev. B **66**, 214412 (2002).

[5] J. Kim, A. H. Said, D. Casa, M. H. Upton, T. Gog, M. Daghofer, G. Jackeli, J. van den Brink, G. Khaliullin, and B. J. Kim, Phys. Rev. Lett. **109**, 157402 (2012).

[6] J. W. Kim, Y. Choi, J. Kim, J. F. Mitchell, G. Jackeli, M. Daghofer, J. van den Brink, G. Khaliullin, and B. J. Kim, Phys. Rev. Lett. **109**, 037204 (2012).

[7] G. Jackeli and G. Khaliullin, Phys. Rev. Lett. **102**, 017205 (2009).

[8] A. Shitade, H. Katsura, J. Kuneš, X. –L. Qi, S.-C. Zhang, and N. Nagaosa, Phys. Rev. Lett. **102**, 256403 (2009).

[9] D. Pesin and L. Balents, Nature Physics **6**, 376 (2010).

[10] J. Chaloupka, G. Jackeli, and G. Khaliullin, Phys. Rev. Lett. **105**, 027204 (2010).

[11] F. Wang and T. Senthil, Phys. Rev. Lett. **106**, 136402 (2011).

[12] Xiangang Wan, A. M. Turner, A. Vishwanath, and S. Y. Savrasov, Phys. Rev. B. **83**, 205101 (2011).

[13] I. I. Mazin, H. O. Jeschke, K. Foyevtsova, R. Valenti, D. I. Khomskii, Phys. Rev. Lett **109**, 197201 (2012).

[14] B. J. Kim, H. Jin, S. J. Moon, J. –Y. Kim, B.-G. Park, C. S. Leem, J. Yu, T. W.

Noh, C. Kim, S. –J. Oh, J. –H. Park, V. Durairaj, G. Cao, and E. Rotenberg, Phys. Rev. Lett. 101, 076402(2008).

[15] P. D. Battle, J. B. Goodenough, R. Price, J. Solid State Chem. **46,** 234 (1983) 234–244.

[16] C. R. Wiebe, J. E. Greedan, and G. M. Luke, Phys. Rev. B **65**, 144413 (2002).

[17] M. A. de Vries, A. C. Mclaughlin, and J.-W. G. Bos, Phys. Rev. Lett. **104**, 177202 (2010).

[18] F. Galasso and W. Darby, Inorg. Chem. **4**, 71 (1965).

[19] G. Blasse, J. Inorg. Nucl. Chem. **27**, 993 (1965).

[20] A. V. Powell, J. G. Gore and P. D. Battle, Journal of Alloys and Compounds **201**, 73 (1993).

[21] R. C. Currie, J. F. Vente, E. Frikkee, and D. J. W. Ijdo, Journal of Solid State Chemistry **116**, 199 (1995).

[22] P. D. Battle and J. G. Gore, J. Mater. Chem. **6**, 1375 (1996).

[23] C. Schinzer and G. Demazeau, J. Mater. Sci. **34**, 251 (1999).

[24] http://elk.sourceforge.net.

[25] T. Aharen, John E. Greedan, Craig A. Bridges, Adam A. Aczel, Jose Rodriguez, Greg MacDougall, Graeme M. Luke, Takashi Imai, Vladimir K. Michaelis, Scott Kroeker, Haidong Zhou, Chris R. Wiebe, and Lachlan M. D. Cranswick, Phys. Rev. B **81**, 224409 (2010).

[26] A. Kismarahardja, J. S. Brooks, A. Kiswandhi, K. Matsubayashi, R. Yamanaka, Y. Uwatoko, J. Whalen, T. Siegrist, and H. D. Zhou, Phys. Rev. Lett. **106**, 056602 (2011).

[27] A. Arrott and J. E. Noakes, Phys. Rev. Lett. 19, 786 (1967).

[28] M. K. Crawford, M. A. Subramanian, R. L. Harlow, J. A. Fernandez-Baca, Z. R. Wang, and D. C. Johnston, Phys. Rev. B **49**, 9198 (1994).

[29] R. J. Cava, B. Batlogg, K. Kiyono, H. Takagi, J. J. Krajewski, W. F. Peck, Jr., L. W. Rupp, Jr., and C. H. Chen, Phys. Rev. B **49**, 11890 (1994).

[30] G. Cao, J. Bolivar, S. McCall, J. E. Crow, and R. P. Guertin, Phys. Rev. B **57**, R11039 (1998).

[31] A. S. Erickson, S. Misra, G. J. Miller, R. R. Gupta, Z. Schlesinger, W. A. Harrison, J. M. Kim, and I. R. Fisher, Phys. Rev. Lett. **99**, 016404 (2007).

[32] K. E. Stitzer, M. D. Smith, and H. zur Loye, Solid State Sci. **4**, 311 (2002).

[33] J. E. Greedan, S. Derakshan, F. Ramezanipour, J. Siewenie and T. Proffen, J. Phys.: Condens. Matter **23**, 164213 (2011).

[34] H. Watanabe, T. Shirakawa, and S. Yunoki, Phys. Rev. Lett. **105**, 216410 (2010).

[35] C. Martins, M. Aichhhorn, L. Vaugier, and S. Biermann, Phys. Rev. Lett. **107**, 266404 (2011).

[36] R. Arita, J. Kuneš, A. V. Kozhevnikov, A. G. Eguiluz, and M. Imada, Phys. Rev. Lett. **108**, 086403 (2012).

[37] A. Subedi, Phys. Rev. B **85**, 020408 (R) (2012).

[38] K. W. Lee and W. E. Pickett, Europhysics Letters **80**, 37008 (2007).

[39] N. A. Bogdanov, V. M. Katukuri, H. Stoll, J. van den Brink, and L. Hozoi, Phys. Rev. B **85**, 235147 (2012).

**Table I:** Structure parameters for La$_2$ZnIrO$_6$[a] and La$_2$MgIrO$_6$[b] obtained from Rietveld refinements of powder XRD data.

| Atom | La$_2$MgIrO$_6$ | | | | La$_2$ZnIrO$_6$ | | | |
|---|---|---|---|---|---|---|---|---|
| | x | y | z | Occupancy | x | Y | z | Occupancy |
| La | 0.4999(1) | 0.5368(7) | 0.2495(6) | 0.997(6) | 0.5019(6) | 0.5469(2) | 0.2499(8) | 0.992(5) |
| Mg/Zn | 0.5 | 0.0 | 0.0 | 1.000(3) | 0.5 | 0.0 | 0.0 | 1.001(2) |
| Ir | 0.0 | 0.5 | 0.0 | 1.000(0) | 0.0 | 0.5 | 0.0 | 0.989(7) |
| O(1) | 0.215(0) | 0.215(5) | -0.045(3) | 1.0 | 0.203(4) | 0.209(2), | -0.0482(2) | 1.0 |
| O(2) | 0.295(0) | 0.704(0) | -0.042(0) | 1.0 | 0.296(1) | 0.700(1), | -0.039(5) | 1.0 |
| O(3) | 0.4214(3) | -0.0180(5) | 0.251(0) | 1.0 | 0.4120(8) | -0.0210(8) | 0.252(1) | 1.0 |

[a] La$_2$MgIrO$_6$: a=5.5883(1) Å, b=5.6284(0) Å, and c= 7.9144(4) Å.

[b] La$_2$ZnIrO$_6$: a=5.5905(2) Å, b=5.6976(3) Å, and c= 7.9344(5) Å.

**Table II:** The $\langle \vec{L} \rangle$ and $\langle \vec{S} \rangle$ expectation values computed over Ir muffin-tin spheres and the band gap $E_{gap}(eV)$ for some values on-site Coulomb repulsion $U$ (eV) and Hund's coupling $J$ (eV). The moments are in units of Bohr magneton.

|  | site | $\langle \vec{L} \rangle$ | $\langle \vec{S} \rangle$ | $\langle \vec{J} \rangle$ | U and $E_{gap}$ |
|---|---|---|---|---|---|
| La$_2$MgIrO$_6$ | Ir(1) | (0.24, -0.22, -0.11) | (0.07, -0.07, -0.02) | (0.31, -0.29, -0.13) | U=2.0, |
|  | Ir(1') | (-0.24, -0.22, 0.11) | (-0.07, -0.07, 0.02) | (-0.31, -0.29, 0.13) | E$_{gap}$=0.41 |
|  | Ir(1) | (0.24, -0.21, -0.11) | (0.07, -0.07, -0.02) | (0.30, -0.28, -0.13) | U=1.5, |
|  | Ir(1') | (-0.24, -0.21, 0.11) | (-0.07, -0.07, 0.02) | (-0.30, -0.28, 0.13) | E$_{gap}$=0.28 |
|  | Ir(1) | (0.23, -0.20, -0.10) | (0.07, -0.07, -0.02) | (0.30, -0.27, -0.13) | U=1.25, |
|  | Ir(1') | (-0.23, -0.20, 0.10) | (-0.07, -0.07, 0.02) | (-0.30, -0.27, 0.13) | E$_{gap}$=0.22 |
|  | Ir(1) | (0.23, -0.20, -0.10) | (0.07, -0.07, -0.02) | (0.30, -0.27, -0.13) | U=1.0 |
|  | Ir(1') | (-0.23, -0.20, 0.10) | (-0.07, -0.07, 0.02) | (-0.29, -0.27, 0.13) | E$_{gap}$=0.16 |
| La$_2$ZnIrO$_6$ | Ir(1) | (-0.18, 0.29, 0.01) | (-0.06, 0.10, 0.00) | (-0.24, 0.39, 0.00) | U=2.0 |
|  | Ir(1') | (0.18, 0.29, -0.01) | (0.06, 0.10, 0.00) | (0.24, 0.39, 0.00) | E$_{gap}$=0.36 |
|  | Ir(1) | (-0.18, 0.29, 0.01) | (-0.06, 0.10, 0.00) | (-0.24, 0.38, -0.01) | U=1.5 |
|  | Ir(1') | (0.18, 0.29, -0.01) | (0.06, 0.10, 0.00) | (0.24, 0.38, -0.01) | E$_{gap}$=0.24 |
|  | Ir(1) | (-0.18, 0.29, 0.01) | (-0.06, 0.10, 0.00) | (-0.24, 0.38, 0.01) | U=1.25 |
|  | Ir(1') | (0.18, 0.29, -0.01) | (0.06, 0.10, 0.00) | (0.24, 0.38, -0.01) | E$_{gap}$=0.18 |
|  | Ir(1) | (-0.18, 0.27, 0.01) | (-0.06, -0.09, 0.00) | (-0.24, 0.37, 0.01) | U=1.0 |
|  | Ir(1') | (0.18, 0.27, -0.01) | (0.06, 0.09, 0.00) | (0.24, 0.37, -0.01) | E$_{gap}$=0.12 |

**Figure captions:**

FIG. 1. (Color online) (a) and (b): Observed, calculated and difference profiles of Rietveld refined room temperature XRD data of powder $La_2MgIrO_6$ and $La_2ZnIrO_6$. Left panels show the crystal structure of $La_2AIrO_6$ for A = Mg/Zn.

FIG. 2. (Color online) (a) and (c): Zero-field-cooled and field-cooled magnetization M(T) at applied field $\mu_0H$ =1 T for $La_2MgIrO_6$ (a) and $La_2ZnIrO_6$ (c). (b) and (d) Inverse susceptibility $\chi^{-1}$ versus temperature T data (solid line) and a fit by a Curie-Weiss model (dashed line).

FIG. 3. (Color online) M(H) curves at T = 2 K for (a) $La_2MgIrO_6$ and (b) $La_2ZnIrO_6$. (c): Modified Arrott plot $(M)^{1/\beta}$ vs $(H/M)^{1/\gamma}$ for $La_2ZnIrO_6$.

FIG. 4. (Color online) heat capacity $C_p$ versus T under 0 and 12 T, respectively for $La_2MgIrO_6$ (a) and $La_2ZnIrO_6$ (c). The difference heat capacity $\Delta C$ and difference entropy $\Delta S$ vs. T data between T=1.9 and 40K for (b) $La_2MgIrO_6$ and (d) $La_2ZnIrO_6$

FIG. 5. (Color online) Neutron scattering data for (a) $La_2MgIrO_6$ and (b) $La_2ZnIrO_6$. The main panels show the temperature dependence of the scattering for $|Q|{\sim}0.79$ Å$^{-1}$. The insets show the scattering above and below $T_N$. The lines are guides to the eye. (c) and (d) Possible magnetic structures predicted from first principles.

FIG. 6. (Color online) (a) and (b) Electrical resistivity ρ as a function of temperature with an applied field of 0 T and 12 T for $La_2AIrO_6$ with A=Mg/Zn. Inset of (a) and (b) give ρ vs 1/T.

FIG. 7. a(1) and b(1): LDA+SO paramagnetic band structure for $La_2MgIrO_6$ and $La_2ZnIrO_6$, respectively, with SOC included via a second variational step. (a2) and (b2): LDA+SO+U band structure for $La_2MgIrO_6$ and $La_2ZnIrO_6$, respectively. Here, U

= 1.0 eV is used and the bands are exchange split only for the LDA+SO+U cases. The Fermi energy is at 0 eV in these plots.

**FIG. 8.** (Color online) (a1) and (b1): Paramagnetic LDA DOS with $t_{2g}$ and $e_g$ projections for $La_2MgIrO_6$ and $La_2ZnIrO_6$, respectively; (a2) and (b2): paramagnetic LDA+SO DOS with $J_{eff}$ projections for $La_2MgIrO_6$ and $La_2ZnIrO_6$.

**FIG. 1 ( GX Cao et al. )**

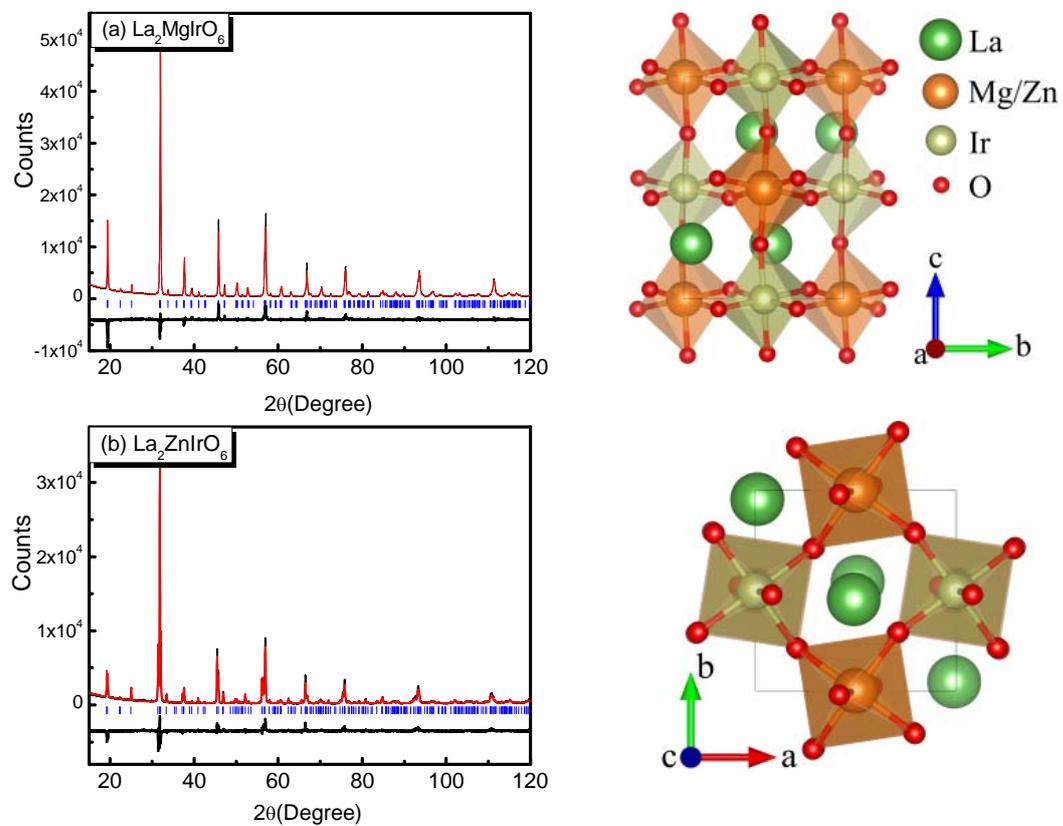

**FIG. 2( GX Cao et al. )**

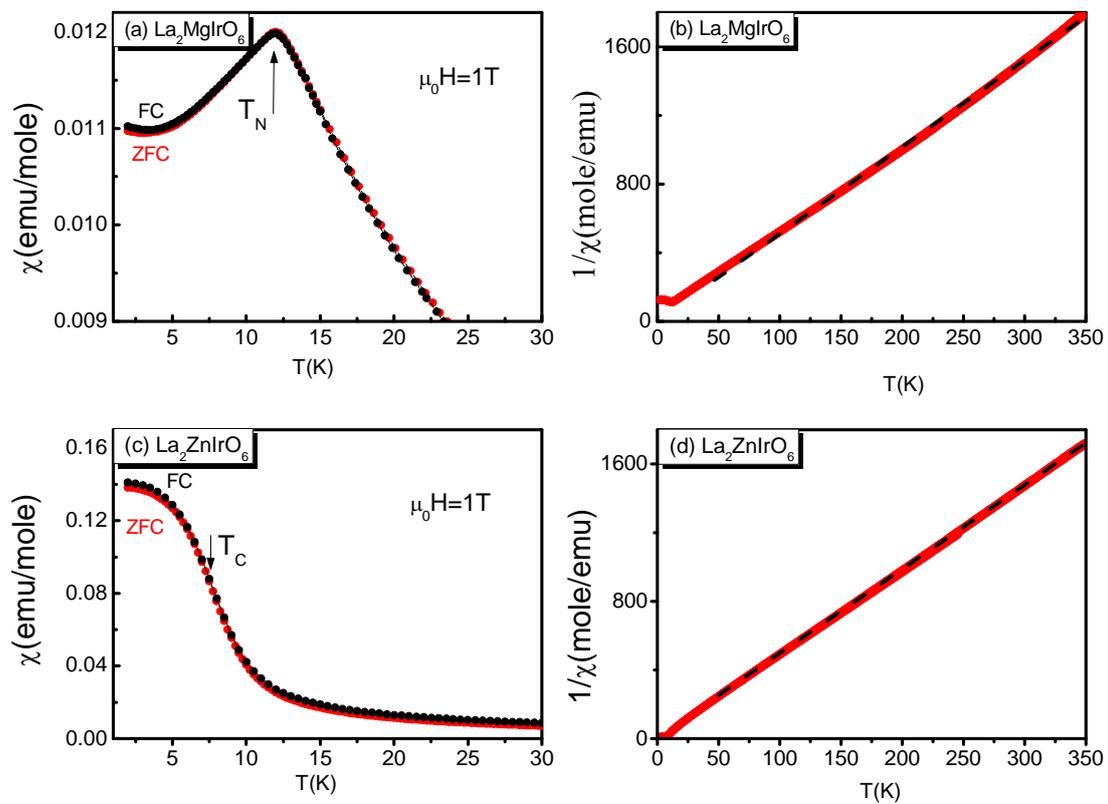

**FIG. 3 ( GX Cao et al. )**

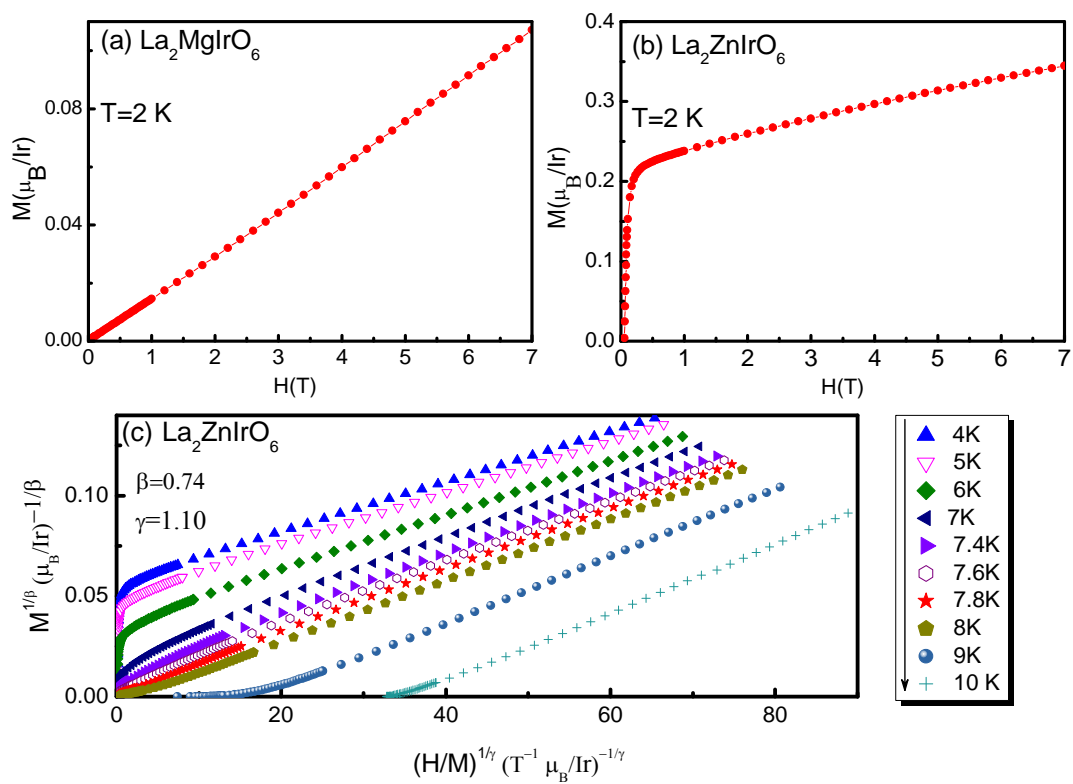

**FIG. 4 ( GX Cao et al. )**

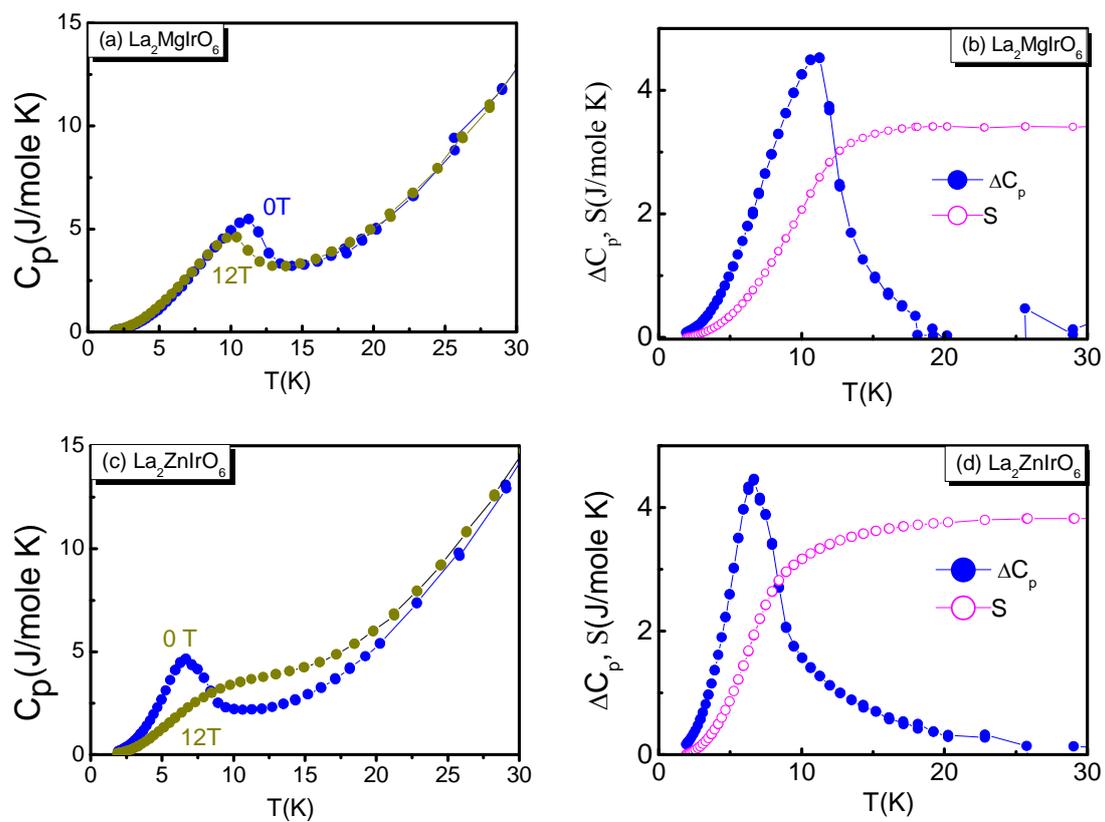

**FIG. 5 ( GX Cao et al. )**

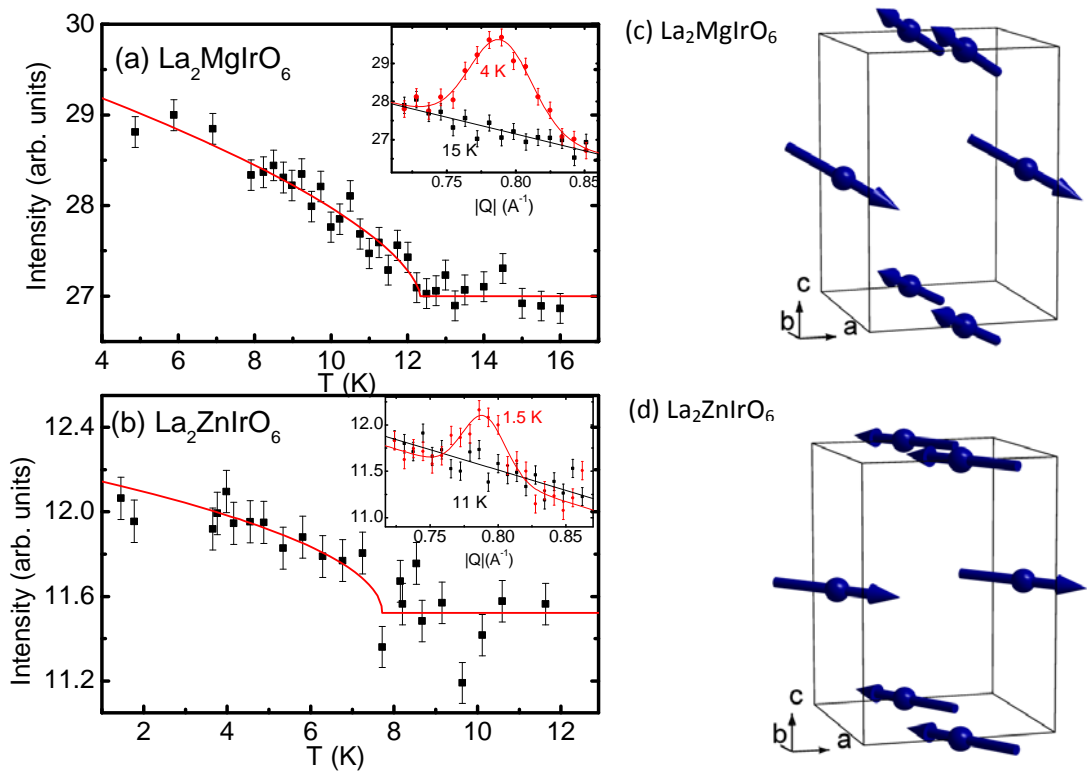

**FIG. 6 ( GX Cao et al.)**

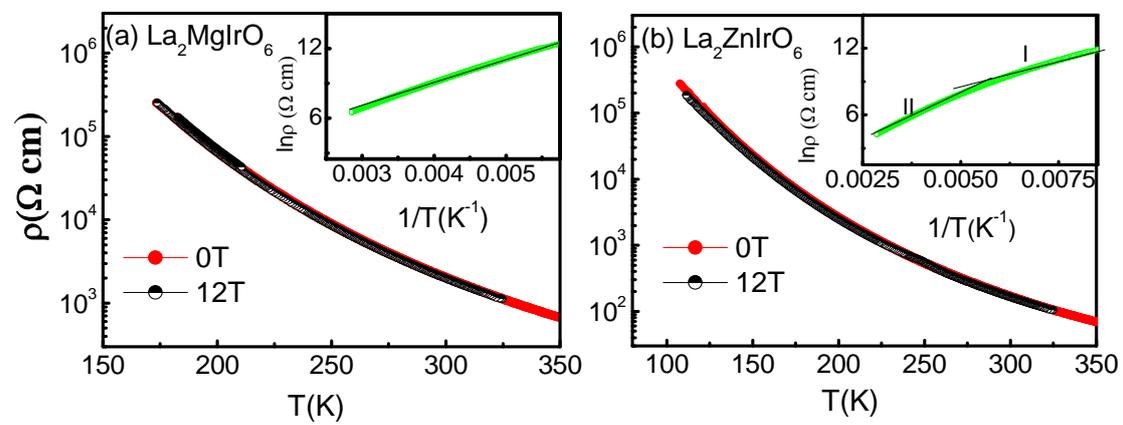

**FIG. 7 ( GX Cao et al.)**

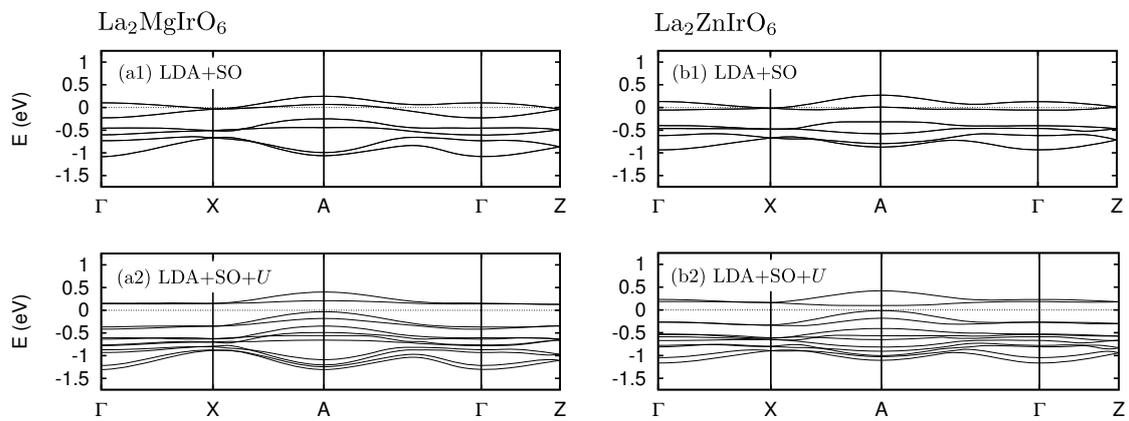

**FIG. 8 ( GX Cao et al. )**

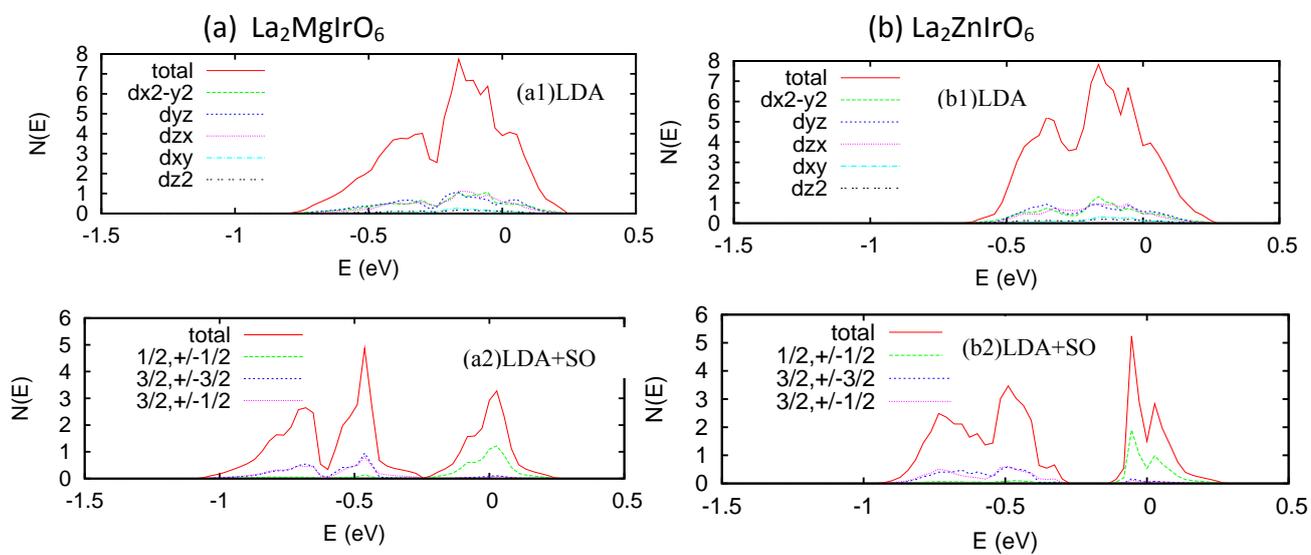